# Laser-accelerated, low divergence 15 MeV quasi-monoenergetic electron bunches at 1 kHz

F. Salehi, M. Le, L. Railing, and H. M. Milchberg

*Institute for Research in Electronics and Applied Physics*
*University of Maryland, College Park, MD 20742*

*Abstract*: We demonstrate laser wakefield acceleration of quasi-monoenergetic electron bunches up to 15 MeV at 1 kHz repetition rate with 2.5 pC charge per bunch and a core with <7 mrad beam divergence. Acceleration is driven by 5 fs, < 2.7 mJ laser pulses incident on a thin, near-critical density hydrogen gas jet. Low beam divergence is attributed to reduced sensitivity to laser carrier envelope phase slip, achieved in two ways using laser polarization and gas jet control: (1) electron injection into the wake on the gas jet's plasma density downramp, and (2) use of circularly polarized drive pulses. Under conditions of mild wavebreaking in the downramp, electron beam profiles have a 2D Lorentzian shape consistent with a $\kappa$ ("kappa") electron energy distribution. Such distributions had previously been observed only in space or dusty plasmas. We attribute this shape to the strongly correlated collisionless bunch confined by the quadratic wakefield bubble potential, where transverse velocity space diffusion is imparted to the bunch by the red-shifted laser field in the bubble.

## I. INTRODUCTION

There is a large demand for high energy electron beams with a wide range of parameters for various applications in science, industry, and medicine [1,2]. Laser wakefield acceleration (LWFA) in plasmas [3,4] could potentially replace conventional accelerators for some of these tasks and open up new applications due to their compact footprint and bright, ultrashort bunches [5]. In the last two decades, advances in peak laser power [6], along with new laser interaction targets [7], have led to the evolution of accelerated electron bunches from wide divergence, tens of MeV thermal spectrum beams [8,9] to ~100 MeV quasi mono-energetic low divergence beams [10–12] and, most recently, quasi mono-energetic bunches up to 8 GeV [13]. The goal of many recent experiments [7,14–16] is achieving up to multi-GeV electron bunch energies in a single acceleration stage; this requires low plasma densities ($N_e/N_{cr} < \sim 10^{-3}$, where $N_{cr}$ is the critical plasma density) to mitigate dephasing, and high laser pulse energy of at least several joules. Consequently, such experiments are limited to low repetition rate ($\leq 10$ Hz) with current laser technology.

For many applications, electron bunch energy in the $1 - 20$ MeV range is of interest, where the main challenges are increasing the repetition rate and the bunch charge, and improving the energy spread, emittance, and laser-to-electron energy conversion efficiency. Early, low repetition rate LWFA experiments applied few-MeV thermal bunches or their bremsstrahlung gamma-rays to radiography [17–20]. More recent ~kHz repetition rate LWFA experiments used the generated ~100 keV beams for electron diffraction [21], and there are proposals to use LWFA sources for electron diffraction at MeV energies [22]. A high repetition rate quasi-monoenergetic multi-MeV electron beam source would benefit all such applications while enabling improved data statistics.

In recent years, multiple groups have generated kHz repetition rate electron bunches using various laser interaction targets. The use of liquid or solid targets has led to large divergence < 3 MeV thermal beams which counter-propagate with respect to the incident laser pulse [23,24]. In experiments using subcritical density gas targets, focusing ~10 mJ laser pulses on the density downramp of an argon or helium jet led to ~100 keV, 10 fC electron bunches [25]. Use of near-critical hydrogen and helium jets by our group enabled relativistic self-focusing of ~10mJ scale laser pulses, and led to the first demonstration of > MeV bunches from gas targets at a kHz repetition rate [26]. However, because our laser pulsewidth was several times the plasma period, these electrons were accelerated in the self-modulated laser wakefield (SM-LWFA) regime, giving a large thermal energy spread with a wide FWHM beam divergence $\theta_{div}$~200 mrad. More recently, tight focusing of 3.4 fs, <2.5 mJ drive pulses on a high density nitrogen gas jet has led to the acceleration of quasi-monoenergetic < 5 MeV, $\theta_{div}$~45 mrad bunches, with electron injection from ionization of He-like nitrogen [27,28].

In this paper, we describe experiments in which we generate up to 15 MeV quasi-monoenergetic low divergence electron bunches using few-cycle, low energy ($\tau$~5 $fs$ FWHM, 2.2-2.7 mJ) laser pulses interacting with near-critical density hydrogen jet targets in the bubble (or blow-out) regime [29,30]. We find that that control of the laser's polarization and focal position in the gas jet can tune the electron beam's energy, divergence, and transverse profile. Under conditions of downramp injection, we measure Lorentzian electron beam profiles corresponding to a so-called 'kappa' electron energy distribution. The profiles are characteristic of a strongly correlated collisionless electron bunch trapped in the bubble quadratic potential and immersed in a laser radiation field [31]. Prior observations of kappa-distributed electrons have been made mainly in space or dusty plasmas [32–35]. Under optimized conditions using circularly polarized drive pulses, we obtain quasi mono-energetic electron beams with energy $E_b$~15 MeV and $\Delta\theta_{div}$ < 7 mrad FWHM divergence. 3D particle-in-cell (PIC) simulations [36] explain the role played in our results by the electron injection location and by laser polarization and carrier envelope phase slip.

## II. EXPERIMENTAL SETUP

The few-cycle, non-carrier envelope phase (CEP) stabilized, LWFA drive pulse is generated by guiding and self-phase modulation (SPM) of a ~35 fs FWHM, < 6 mJ, λ = 800 nm Ti:Sapphire laser pulses in a 2.5 m long, 500 μm inner diameter hollow core fiber (HCF) with helium gas injected near the fiber exit and pumped out near the entrance [37]. See Fig. 1(a). Before injection into the HCF, the pulse polarization is adjusted by a quarterwave-plate, where, depending on the input ellipticity, the HCF exit spectrum (Fig. 1(b)) can reach ~200 nm FWHM with a central wavelength $\lambda_0$~650 nm. After propagation through a chirped mirror compressor (90% throughput) and a pair of wedges for fine tuning the pulse length [38], the beam is directed into the experimental chamber. The pulse loses more than half of its energy through leakage from the HCF and losses from the compressor, beam routing mirrors, and windows. Ultimately, 2.2 mJ − 2.7mJ in a 5 fs FWHM pulse [37] is focused by an $f/6.5$ off-axis paraboloid (OAP) to a 4.5 μm FWHM intensity spot on the target, yielding a peak vacuum intensity of $2.3 \times 10^{18}$ W/cm$^2$ (peak normalized vector potential $a_0 = 0.9$). A probe pulse split from the compressed pulse is used for interferometry and shadowgraphy of the gas jet target and the laser-target interaction [37]. The near-critical density jet is produced by feeding high pressure hydrogen gas through a solenoid valve to a Mach 2.9 supersonic nozzle with a 100μm diameter throat. The near-Gaussian gas density profile, measured with the probe pulse, is shown in Fig. 1(c). In the laser beam path ~100 μm above the nozzle orifice, the jet is ~150 μm FWHM with a peak H$_2$ density adjusted in the range $N_m = 1.0 - 1.7 \times 10^{20}$ cm$^{-3}$. When fully ionized, this gives electron density in the range $N_e/N_{cr} = 0.08 - 0.13$, where $N_{cr} = 2.64 \times 10^{21}$ cm$^{-3}$ is the critical plasma density at $\lambda_0 = 650$ nm. Earlier versions of this gas jet [39] were used to generate near-critical density targets for experiments in the SM-LWFA regime [26,40,41]. For the conditions of this experiment, we observed electron beams only from hydrogen jets. No comparable electron beams were observed using other gases (He, N$_2$, Ar), for which PIC simulations show that ionization-induced defocusing refracts and distorts the pulse before the onset of relativistic self-focusing. Tighter focusing at $f/3.2$ partially mitigated ionization defocusing, but led to electron beams with lower energy and higher divergence [38].

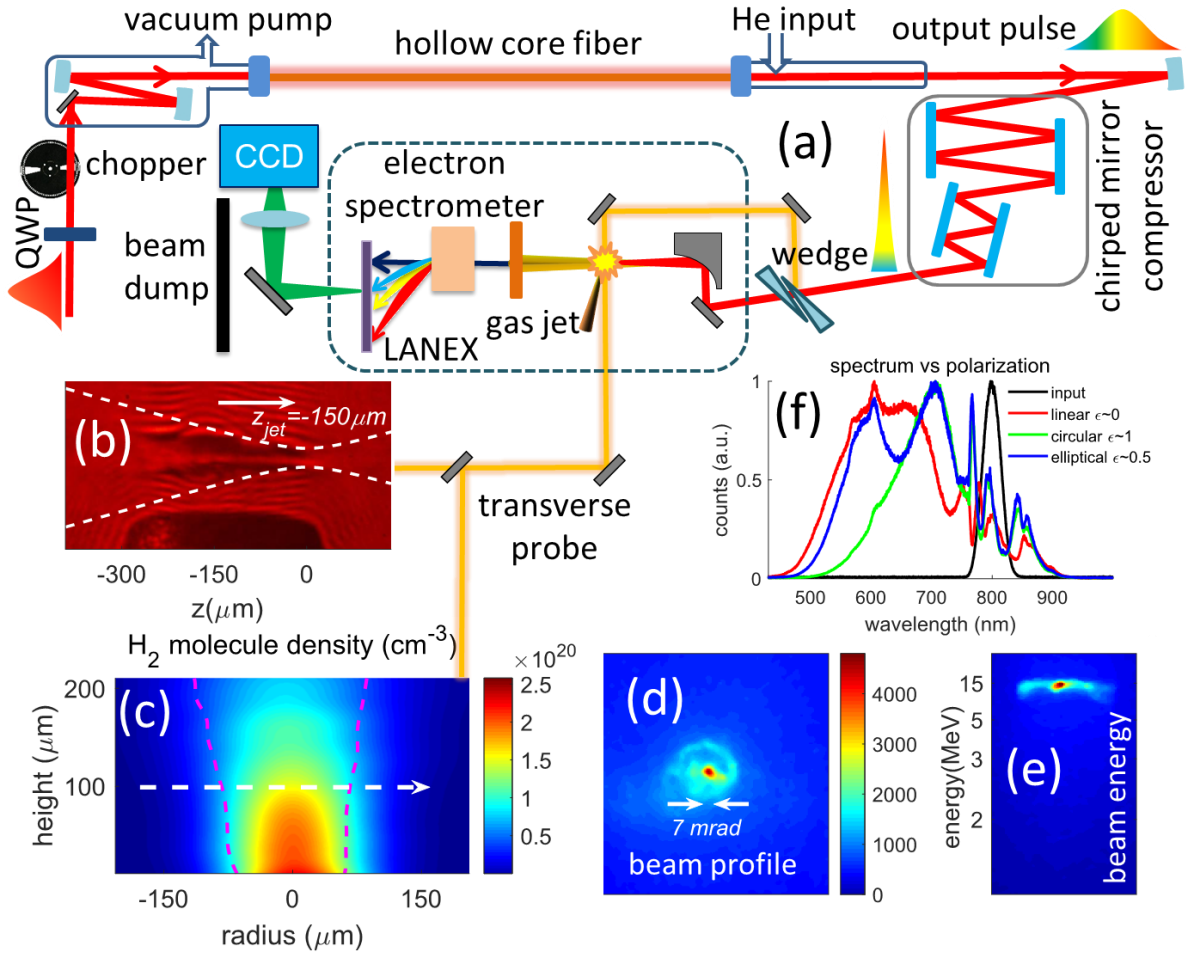

**Figure 1.** Experimental setup. **(a)** Laser pulses with few-cycle duration, generated through a hollow core fiber (HCF) and a chirped mirror compressor, are used to drive electron acceleration from a near-critical density hydrogen jet target. A probe pulse split from the main drive pulse is used for **(b)** probing of the few-cycle pulse interaction with the jet and **(c)** interferometric measurement of the jet density. The white arrows show the drive beam propagation direction, the white dashed lines ((b)) indicate the laser beam $4e^{-2}$ Gaussian intensity envelope, and the red dashed lines ((c)) show the density FWHM contour. **(d,e)** Sample electron beam profile and energy spectrum as imaged on the LANEX screen. **(f)** HCF output spectrum as a function of input polarization (input energy 6 mJ). Elliptical input polarization generates the broadest output spectrum, which is near-circularly polarized.

For electron energy spectrum measurements, three different sets of permanent magnet dipoles with magnetic field $\sim 0.08T - 0.35T$ were used between the jet and a LANEX fluorescing screen (located ~30cm beyond the jet), which was imaged by a low noise CCD camera. Full electron beam profiles were measured by translating the magnet out of the beam path. Figure 1 shows the magnetic spectrometer (Fig. 1(a)) and examples of an electron beam profile and energy spectrum (Fig. 1(d) and 1(e)).

The radiation dose from the electron beam measured in the forward direction, from bremsstrahlung conversion in the aluminum and lead beam dump, is $> 1$ µRad/shot for the highest energy electron bunches ($E_b \sim 15$ MeV) measured in the experiment. To avoid high dose accumulation at the kHz pulse repetition rate of the laser, we opened the gas jet's solenoid valve for a $\sim 10$ ms interval every 2 s. In addition, when generating $> 10$ MeV electron bunches, we used a chopper in the laser beam path to lower the pulse repetition rate to 100Hz. This lowered the accelerator average repetition rate to 0.5 Hz and reduced the accumulated dose by 2000 ×. We expect that removing the chopper and/or the solenoid valve would result in the same electron beam at a higher repetition rate, with minor changes due to the small pressure buildup in the vacuum chamber, as observed in our prior SM-LWFA experiment [26].

The maximum pulse energy injected into the HCF is limited by the critical self-focusing power and the ionization threshold of the gas flowing in the fiber, here helium. Ionization of He leads to significant blue shifting of the spectrum [42], increased coupling and guiding losses, and deterioration of the spectral phase, making compression to few-cycle pulsewidths difficult. The highest input pulse energy is set near the threshold at which filamentation in He is observed at the fiber entrance. Circularly polarized (CP) pulses have a higher tunneling ionization threshold than linear polarized (LP) pulses, so it is typical to inject gas-filled hollow core fibers with higher energy CP pulses, generate a SPM-broadened spectrum, and then convert the fiber output back to LP [43,44]. In our experiment, we observe that the use of elliptically polarized input pulses leads to larger bandwidth and, in turn, shorter compressed pulses than with LP or CP input pulses, where partial He ionization near the HCF entrance plays a role [45]. Figure 1(b) shows fiber exit spectra for a 6 mJ input pulse for a range of polarization ellipticities $\varepsilon$, the ratio of minimum to maximum electric field in the polarization ellipse. The largest output bandwidth occurs for $\varepsilon \sim 0.5$, where the output pulse has a double hump spectrum with $\sim 200$ nm bandwidth. In this case, the ellipticity evolves during the pulse propagation in the fiber [46,47], and we measure nearly CP output pulses, $\varepsilon \sim 0.9$ [45]. Here, the shortest pulse duration after compression is 5 fs. For CP input pulses, the comparatively smaller exit bandwidth precludes compression below $\sim 7$ fs, and no significant electron acceleration was observed with those beams.

## III. RESULTS AND DISCUSSION

Figure 2(a) shows profiles of accelerated electron bunches as a function of the jet center position $z = z_{jet}$ with respect to the laser focus ($z = 0$). Here, we used 2.2 mJ, 5 fs LP drive pulses, and the peak electron density at jet center was $N_{e,peak}/N_{cr} = 0.08$. This gives $(P/P_{cr})_{max} \sim 2$ for these conditions, where the minimum critical power for relativistic self-focusing is $P_{cr,min} = 17.4(N_{cr}/N_{e,peak})$ GW [48]. Images are 10 shot averages; these electron beams were quite stable, with shot-to-shot pointing jitter as small as 2 mrad. Starting at $z_{jet} = -200$µm (vacuum focus 200µm beyond the jet center), we observed a barely detectable electron beam ($\sim 10$ fC) with a large divergence angle ($\Delta\theta_{div} > 50$ mrad) and low energy ($E_b \gtrsim 0.2$ MeV). Moving the jet forward toward the focus, $\Delta\theta_{div}$ continuously decreases to minimum of 20 mrad (Fig. 2(a), $z_{jet} = -80$ μm), where the laser waist is located near the jet's far side half-maximum density, generating ~2 pC, ~0.5 MeV quasi-monoenergetic (QME) beams. Moving the jet center closer to the beam waist, the divergence increases again to $\Delta\theta_{div}$ ~34 mrad (Fig. 2(a), $z_{jet} = -40$μm). As the jet is moved to the beam waist at $z_{jet} = 0$, $\Delta\theta_{div}$ and $E_b$ continuously increase to >70 mrad and ~3 MeV QME beams, but the total charge per shot drops and the beam disappears as the jet center is moved past the beam waist ($z_{jet} = +20$ μm).

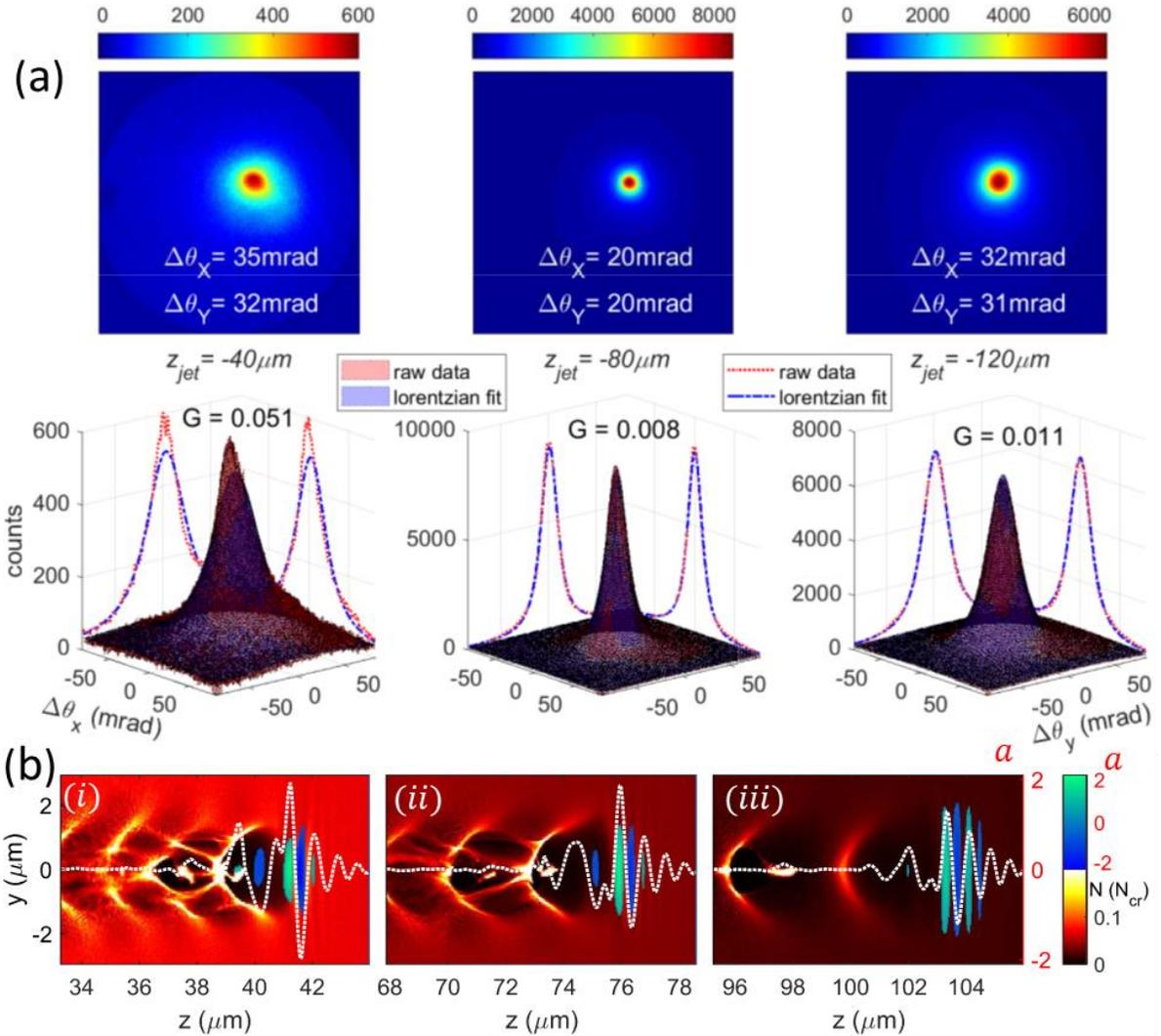

**Figure 2.** 2.2 mJ, 5 fs LP drive pulse, peak jet plasma density $N_{e,peak}/N_{cr} = 0.08$, $(P/P_{cr})_{max}$~2. (a) Electron beam profiles (upper panels) and 2D Lorentzian fits (lower panels) for varying position of the jet center ($z_{jet}$) relative to the laser beam waist at $z = 0$ (integrating over 10 shots). The Lorentzian fit quality is reduced as the jet center moves closer to the beam waist, with RMS fit error G [37] shown in each panel. (b) PIC simulations, showing normalized laser vector potential $a$ (colourmap and dashed white line) overlaid on electron density. Propagation is launched at beam waist located (i) 10 μm past jet center, (ii) 30 μm past jet center, and (iii) 50 μm past jet center, progressively farther into the density downramp. The horizontal scale is distance from jet center. Simulation parameters: $a_0 = 1$, 5 fs LP drive pulse, peak plasma density $N_{e,peak}/N_{cr} = 0.08$, jet FWHM 140 μm.

The simulations of Fig. 2(b) qualitatively explain the trends as follows: When electron injection occurs in the density downramp [49,50] (the half-peak density is at $z = 70$μm), the laser spot $w_l$ matches the bubble radius, $w_l \sim R_b = 2k_p^{-1}\sqrt{a} \sim 1.4$ μm [30], where $k_p$ is the plasma wavenumber and $a \sim 1.8$ is the peak normalized laser vector potential in the plasma. Here, the adiabatic expansion of the bunch in the downramp reduces its divergence [51,52] (panel $(ii)$). Further in the downramp (panel $(iii)$), injection occurs only in the lower phase velocity second plasma wave bucket, where the defocusing force behind the first bucket imposes a larger beam divergence. Closer to the jet density peak, self-injection from wavebreaking becomes important (panel $(i)$), which degrades bunch divergence.

The electron beam profiles shown in Fig. 2(a) are an excellent fit to the 2D Lorentzian profile $\sigma_q \propto [1 + ((x-x_0)/w_x)^2 + ((y-y_0)/w_y)^2]^{-1.5}$, where $\sigma_q(x,y)$ is the charge density on the screen, $(x_0, y_0)$ is the profile center, and $w_x, w_y$ are the charge spread in each dimension. For the images shown here, LANEX fluorescence is linear in the local beam flux [37]. The goodness of fit is quantified by the normalized RMS error G [37] indicated in Fig. 2(a), where it seen that G deteriorates as the beam waist moves closer to the jet center or past the midrange of the downramp. We observe the Lorentzian charge density profile for almost any pulse energy and plasma density if the jet center is located far enough upstream of the vacuum focus. We note that the measured beam profiles preserve the transverse bunch structure exiting the plasma. While the effect of space charge on bunch emittance and distortion can be considerable for non-relativistic electron beams, it is negligible for the bunch energies of our experiments. Our PIC simulations and those of other groups [51–53] confirm that the transverse bunch profile grows linearly after exiting the plasma, and the beam divergence remains constant.

The Lorentzian beam profiles in Fig. 2 are consistent with a "kappa" distribution [33,54–57], which is often used in space physics to describe a quasi-stationary energy distribution of strongly correlated collisionless electrons far from equilibrium. Under a kappa distribution, after integrating over the velocity degrees of freedom, the spatial number density distribution of an electron bunch in the presence of a transverse potential $\Phi_\perp(r)$ takes the form $N(\Phi_\perp(r)) \propto (1 + (\kappa - 3/2)^{-1} q\Phi_\perp(r)/k_B T_\perp)^{-\kappa+1/2}$, where $q = -e$, and kappa can take on values $3/2 < \kappa < \infty$. Here, $\kappa \to 3/2$ describes collisionless plasmas with the highest possible correlations, and $\kappa \to \infty$ gives $\lim_{\kappa \to \infty}\{N(\Phi_\perp)\} \propto \exp(-e\Phi_\perp/k_B T_\perp)$, converging to the Maxwellian distribution of uncorrelated electrons. In our case, $\Phi_\perp(r) \propto r^2$ is the quadratic electrostatic potential of the wakefield bubble [4]. This leads to a 2D Lorentzian profile, with $\kappa = 2$ giving the best fit to our measured $\sigma_q(x,y)$ transverse profiles. The strong correlations stem from the extremely compact space-time volume of the electron bunch made possible by the few-cycle drive pulse, the near-critical density target, and the presence of red-shifted laser field [58,59] extending back through the second bucket (Fig. 2(b)). The PIC simulations of Fig. 2(b) give bunch length $l_{bunch} \sim 250$ nm (0.8 fs), bunch width $d_{bunch} \sim 350 nm$, and density $N_{bunch} \sim 5 \times 10^{21}$ cm$^{-3}$. A bunch transverse temperature $k_B T_\perp < 1$ keV is estimated from $\Delta\theta_{div} \sim (k_B T_\perp/mc^2)^{1/2} \sim 20 - 40$ mrad in Fig. 2(a). Under these conditions, the transverse Debye length $\lambda_D \sim 3$ nm is much less than $d_{bunch}$, $N_{bunch}\lambda_D^3 \gg 1$, and the electron-electron collision time is $\tau_{ee} \sim 16\pi N_{bunch}\lambda_D^3/\omega_p \sim 10$ ps, much longer than the ~100 fs bunch transit time through the plasma. A physical mechanism setting the specific value of $\kappa$ is demonstrated by Hasegawa *et al.* [31] to be transverse velocity space diffusion from coulomb fluctuations induced by the laser field. There, it is shown that $\kappa \sim 0.2(\lambda_D/r_q)^2(1 - N_e/N_{cr})^2 \sim 0.2(\lambda_D/r_q)^2$ where $r_q$ is the electron quiver amplitude in a laser field of amplitude $a_q$. Here, using $N_e = N_{bunch} \sim 2N_{cr}$ and $\kappa = 2$ gives $r_q/\lambda_D \sim 0.32$, from which $a_q/a_{peak} \sim 0.01$. This agrees reasonably well with the simulation in the two rightmost panels of Fig. 2(b), where $a_q/a_{peak} < 0.05$.

Figure 3 shows the electron bunch profile and energy spectrum for higher energy LP drive pulses (2.6 mJ) and higher peak plasma density ($N_{e,peak}/N_{cr} \sim 0.12$), giving $(P/P_{cr})_{max} \sim 3.5$, where we compare results for the vacuum beam waist positioned near jet center ($z_{jet} = 20$ μm) and in the density downramp ($z_{jet} = -110$ μm). As in the experiments of Fig. 2, the electron bunch has higher energy and a larger divergence for the vacuum beam waist near jet center (here $E_b \sim 9$ MeV, $\Delta\theta_{div} \sim 300$ mrad, and charge $\sim 1.7$ pC) with the energy dropping for the beam waist on either side of it. Here, simulations show that the beam's wide divergence and non-Lorentzian profile ($G = 0.14$) result from self-injection via catastrophic wavebreaking. The best pointing stability and lowest divergence occurs for the beam waist in the density downramp, where for $z_{jet} = -110$ μm, we measure $E_b \sim 3$ MeV, $\Delta\theta_{div} \sim 45$ mrad, and accelerated charge $\sim 2.2$ pC per shot. Here, the beam profile is better-fit by a Lorentzian ($G = 0.05$), but $\Delta\theta_{div}$ is larger than in Fig. 2's case of low pulse energy and low plasma density.

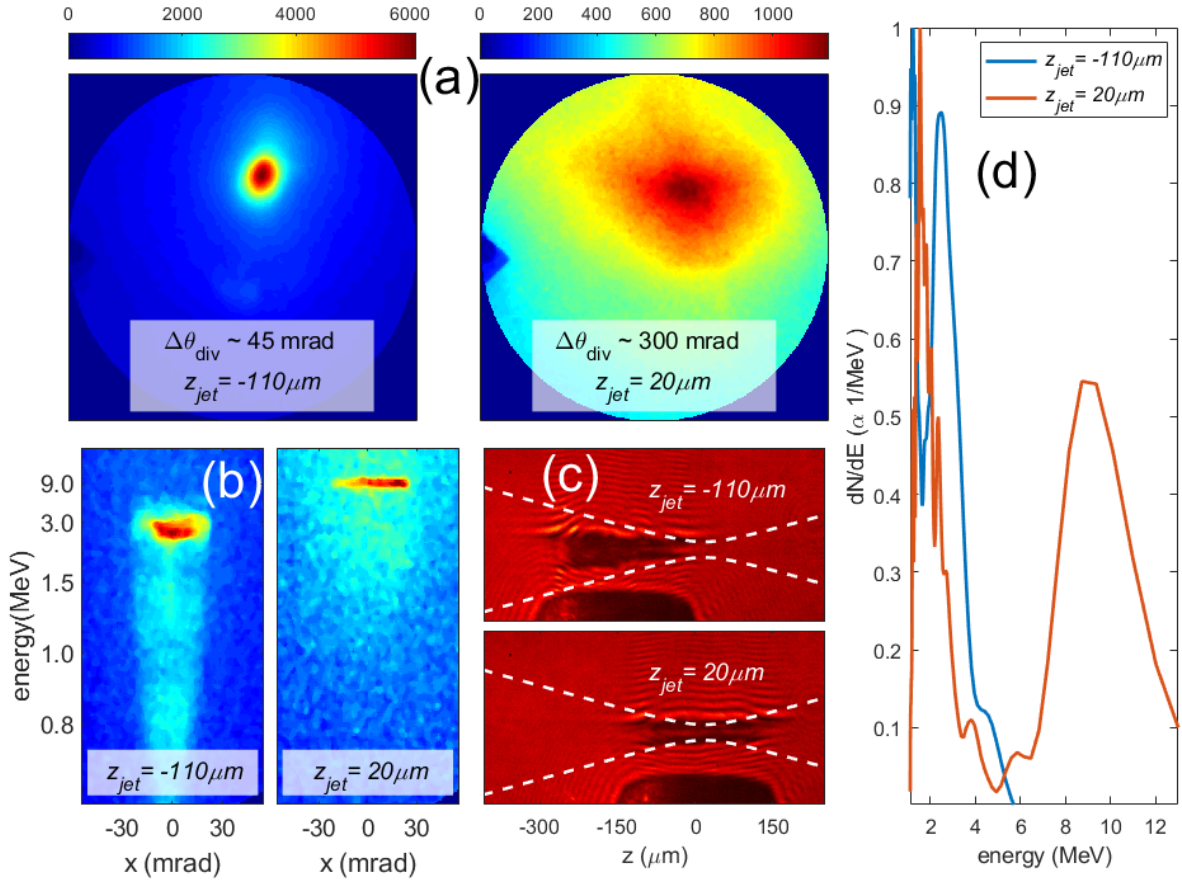

**Figure 3.** 2.6 mJ, 5 fs LP drive pulse, peak jet plasma density $N_{e,peak}/N_{cr} = 0.12$, $(P/P_{cr})_{max} \sim 3.5$. **(a)** Electron beam profiles for the vacuum beam waist near jet center ($z_{jet} = 20$ μm) and in the density downramp ($z_{jet} = -110$ μm) **(b)** Corresponding electron bunch spectra and **(c)** shadowgraphic images, which show the beam focus near the jet center (bottom) and in the downramp (top). The dashed white lines show the $4e^{-2}$ envelope of the vacuum beam. **(d)** Lineouts of electron spectra of (b) on linear energy scale. The resolution of the QME peaks is reduced by some penetration of high energy electrons through the spectrometer slit edges. The electron beam profiles and spectra are 10 shot averages.

Figure 4 shows the result of using a circularly polarized (CP) 5 fs drive pulse, for laser energy 2.7 mJ, $N_{e,peak}/N_{cr} = 0.10$, and $(P/P_{cr})_{max} \sim 3.1$. As in the prior figures, we compare acceleration for varying vacuum beam waist position with respect to the jet center. Starting with the jet center far upstream of the focus, $z_{jet} = -130$ μm (or focus in the far downramp), the beam has a 2D Lorentzian shape (G=0.03), with a divergence significantly reduced to $\Delta\theta_{div} \sim 15$ mrad, $\sim 3\times$ smaller than the LP case. The bunch charge is $\sim 9$ pC and $E_b \lesssim 5$ MeV in a non-QME spectrum. For $z_{jet} = -60$ μm, the $\sim 2.8$ pC beam has a tight central spot ($\Delta\theta_{div} \sim 10$ mrad) with a $\sim 9$ MeV QME peak with $\Delta E_b/E_b \sim 0.2$, surrounded by a lower energy ring structure with a full width divergence $\sim 70$ mrad. The beam image is shown enhanced by $5\times$ in the inset.

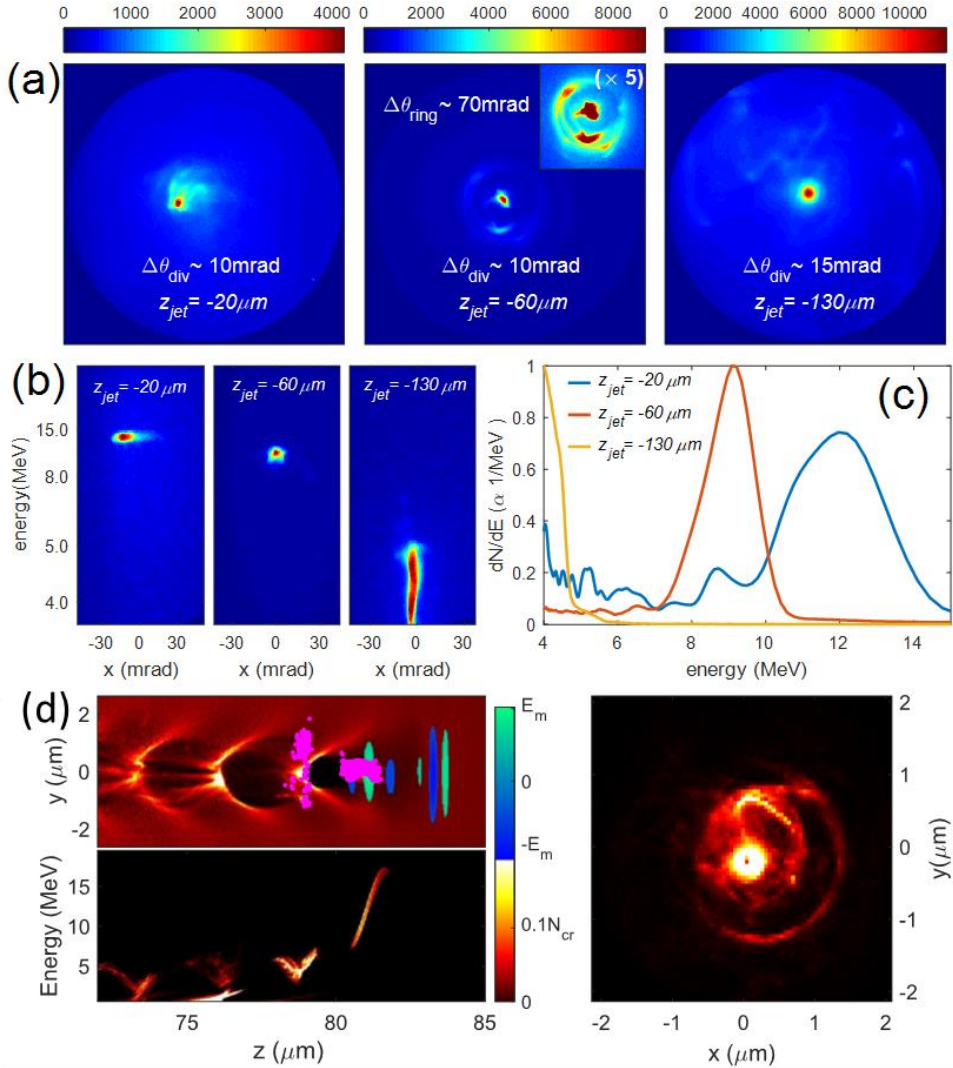

**Figure 4.** 2.7 mJ, 5 fs circularly polarized (CP) drive pulse, peak jet plasma density $N_{e,peak}/N_{cr} = 0.10$, $(P/P_{cr})_{max} \sim 3.1$. **(a)** Single shot electron beam profiles for varying $z_{jet}$. The middle panel inset for $z_{jet} = -60$ μm (near half-peak density in downramp) shows the electron ring image enhanced by 5×. **(b)** Corresponding energy spectra. **(c)** Energy spectra lineouts plotted on a linear scale. **(d)** PIC simulation result for a CP pulse with injection near half-maximum density in downramp. Electrons (magenta dots) injected in the second bucket form a ring in the transverse plane (right panel) as they outrun the first bucket. The electrons in the ring have a lower energy than those in the central spot (bottom left panel). Simulation parameters: pulse energy 2.7 mJ, pulsewidth 5 fs, vacuum spot FWHM 4.5μm, jet FWHM 140 μm, $z_{jet} = -60$ μm, $N_e/N_{cr} = 0.1$.

The ring structure appears only with the CP driver. Moving the laser beam waist further toward the jet center, $z_{jet} = -20$ μm, the ring transforms into a ~50 mrad pedestal surrounding a central spot with $\Delta\theta_{div}$~10 mrad. Here, the bunch has a ~12 MeV QME peak with $\Delta E_b/E_b$ ~0.25 and a total charge~3.5 pC. The divergence of the central peak is ~30× smaller than in the LP case.

We attribute this dramatic reduction in electron beam divergence for a CP driver to reduced laser-induced distortion of the first LWFA bucket (bubble) following the laser pulse. As has been shown in prior simulations, a few-cycle LP driver pulse oscillates the bubble along the laser polarization direction [60,61], distorting the bubble's accelerating field and the accelerated bunch itself. This effect is quite different from the case of hosing instability with many-cycle pulses, where the ponderomotive potential driving the wakefield is similar for LP and CP pulses. The few-cycle effect is enhanced by the large laser red shift and the plasma's negative group velocity dispersion, which causes the laser field to slip to the back of the first bucket and even into the second bucket.

To visualize the effects of a few-cycle driver on the bubble and its accelerating field, Fig. 5 shows the results of simulations contrasting LP and CP pulse interaction with a neutral H$_2$ jet, with the beam waist at jet center, and using our experimental parameters (see figure caption). Figure 5(a) plots the time evolution of the normalized bubble centroid position $(\bar{x}/R_b, \bar{y}/R_b)$ in the transverse plane. Here $(\bar{x}, \bar{y}) = \bar{\mathbf{r}}_\perp$ and $R_b$ are given by $\{\bar{\mathbf{r}}_\perp, R_b\} = \int d\xi d^2\mathbf{r}_\perp \{\mathbf{r}_\perp, |\mathbf{r}_\perp|\} N_e(\mathbf{r}_\perp, \xi)[\int d\xi d^2\mathbf{r}_\perp N_e(\mathbf{r}_\perp, \xi)]^{-1}$, where $R_b$ is the mean bubble radius, $\xi$ is the local position coordinate along the propagation direction in the PIC simulation's moving window, and the integration limits encompass the full volume of the leading wakefield bucket [37]. It is seen that the LP-induced bubble centroid dominantly oscillates along the laser polarization ($y$) direction, with negligible perturbation along $x$, while for CP, the centroid follows a relatively tighter spiral. The centroid excursion is always significantly less for CP, with $|\bar{y}|_{CP,max}/|\bar{y}|_{LP,max} \lesssim 0.3$. In another view, Fig. 5(b) plots $\bar{x}/R_b$ and $\bar{y}/R_b$ vs. time, showing the $\pi/2$ out-of-phase $x$ and $y$ oscillations induced by CP.

The bubble centroid oscillation is caused by the slip of the laser pulse envelope with respect to the laser field oscillations [60,61] —the carrier envelope phase (CEP) slip—and manifests as a bubble modulation frequency $\Delta\omega_m/\omega_0 \sim (v_g - v_p)/c$, where $\omega_0$ is the laser central frequency, and $v_g$ and $v_p$ are the group and phase velocities in the plasma. The CEP slip gives a bubble modulation period $T_{bubb} = 2\pi/\Delta\omega_m \sim (N_{cr}/N_e)T_{laser}$, where $T_{laser} = 2\pi/\omega_0$ is the laser period. The modulation period in Fig. 5(b) is $\omega_0 T_{bubb}$~60, giving $T_{bubb}$~$9.3 T_{laser}$, in good agreement with the density used in the simulation, $N_{cr}/N_{e,peak}$ ~9.1.

The more confined CP-induced spiral motion of the bubble centroid (Fig. 5(a)) is suggestive of a more symmetric accelerating structure compared to the LP case, where the bubble is driven primarily along one transverse axis. To see the effect of transverse bubble motion on bunch acceleration, we plot in Fig. 5(c) the average axial field $\overline{E_z(0,0,\xi)}$ in the accelerating phase of the propagating bubble as a function of time, where the average is taken from maximal negative $E_z$ to $E_z = 0$. The open circles denote the times in the bubble evolution after the onset of bunch injection and acceleration. The LP-induced bubble oscillations clearly result in significantly more $\overline{E_z}$ modulation, which is coupled to greater modulation in the bubble's focusing field $\mathbf{E}_\perp$ and affects the perpendicular bunch momentum $\mathbf{p}_\perp$. Shown in the top panels of Fig. 5(d) are electron bunch transverse profiles at $\omega_0 t = 2000$ ($z = 80$μm) for energy > 2 MeV; the panels below show corresponding $x$ and $y$ lineouts. For CP, the simulation shows a tight central feature surrounded by an electron ring, while the LP result shows a beam elongated along the laser polarization direction ($y$). The divergence angles $\Delta\theta_x$ and $\Delta\theta_y$ are determined from a sequence of beam snapshots, which give the values $\Delta\theta_{div}^{CP} = 9$ mrad and $\Delta\theta_{div}^{LP} > 60$ mrad calculated from $\Delta\theta_{div} = ((\Delta\theta_x^2 + \Delta\theta_y^2)/2)^{1/2}$. These are consistent with the experiment. We note that for LP as opposed to CP, laser-driven electron loss from the simulation window renders $\Delta\theta_{div}^{LP}$ an underestimate. The associated movies of transverse beam evolution under CP and LP are shown in ref. [37], where the ring beams in the CP case are seen to result from spiral motion of accelerated electrons.

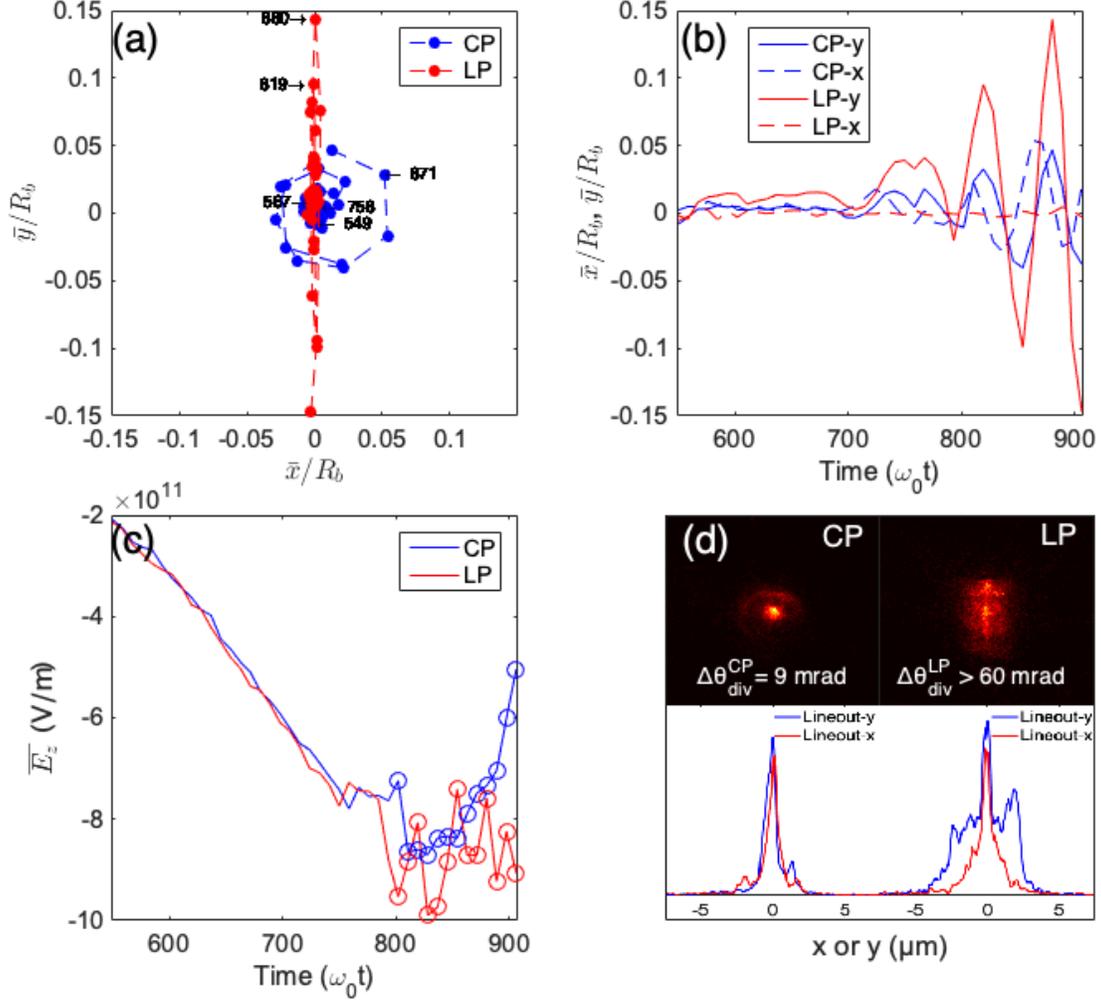

**Figure 5.** Simulation parameters: LP or CP, 2.7 mJ, 5 fs, $\lambda_0 = 650$ nm, vacuum spot FWHM 4.5μm, jet 140 μm FWHM, $N_e/N_{cr} = 0.11$, center of neutral hydrogen jet placed at beam waist ($z_{jet} = 0$). **(a)** Normalized bubble centroid position ($\bar{x}/R_b$, $\bar{y}/R_b$) in transverse plane vs. time. CP=circular polarization, LP=linear polarization. Numbers near points are values of $\omega_0 t$. **(b)** $\bar{x}/R_b$ and, $\bar{y}/R_b$ vs time, showing the CEP slip-induced bubble centroid oscillations (see text). **(c)** Average axial field $\overline{E_z(0,0,\xi)}$ in the accelerating phase of the bubble vs. time for CP and LP pulses. The open circles are for times after electron injection and before bubble breakup. **(d)** Top panels: Electron beam profiles (>2 MeV) at $\omega_0 t = 2000$ from CP and LP pulses, where the bunch is near $z = 80\ \mu m$. The RMS beam divergence $\Delta\theta_{div} = ((\Delta\theta_x^2 + \Delta\theta_y^2)/2)^{1/2}$ is shown in the panels. Bottom panels: beam lineouts along $x$ and $y$.

One scenario for electron ring formation in Fig. 4(a) is shown in the particle tracking PIC simulation of Fig. 4(d), where density downramp-injected electrons in the second bucket are driven in spiral orbits by the delayed and redshifted portion of the CP laser pulse, outrun the back of the first bucket, lose energy by dephasing, and are driven out transversely. The initial symmetry of the secondary bunch driven by CP is responsible for the ring shape, as seen on the right panel of Fig. 4(d). The electrons in the ring have a lower energy than those in the central spot (bottom left panel). For similar downramp focusing in the LP case of Fig. 3, electrons are also injected in the two leading buckets, but the second bucket electrons dephase and quickly diverge in the polarization direction. Previous experiments and simulations using circularly polarized laser pulses, some producing ring-like electron beams, have been reported under various conditions [62–65].

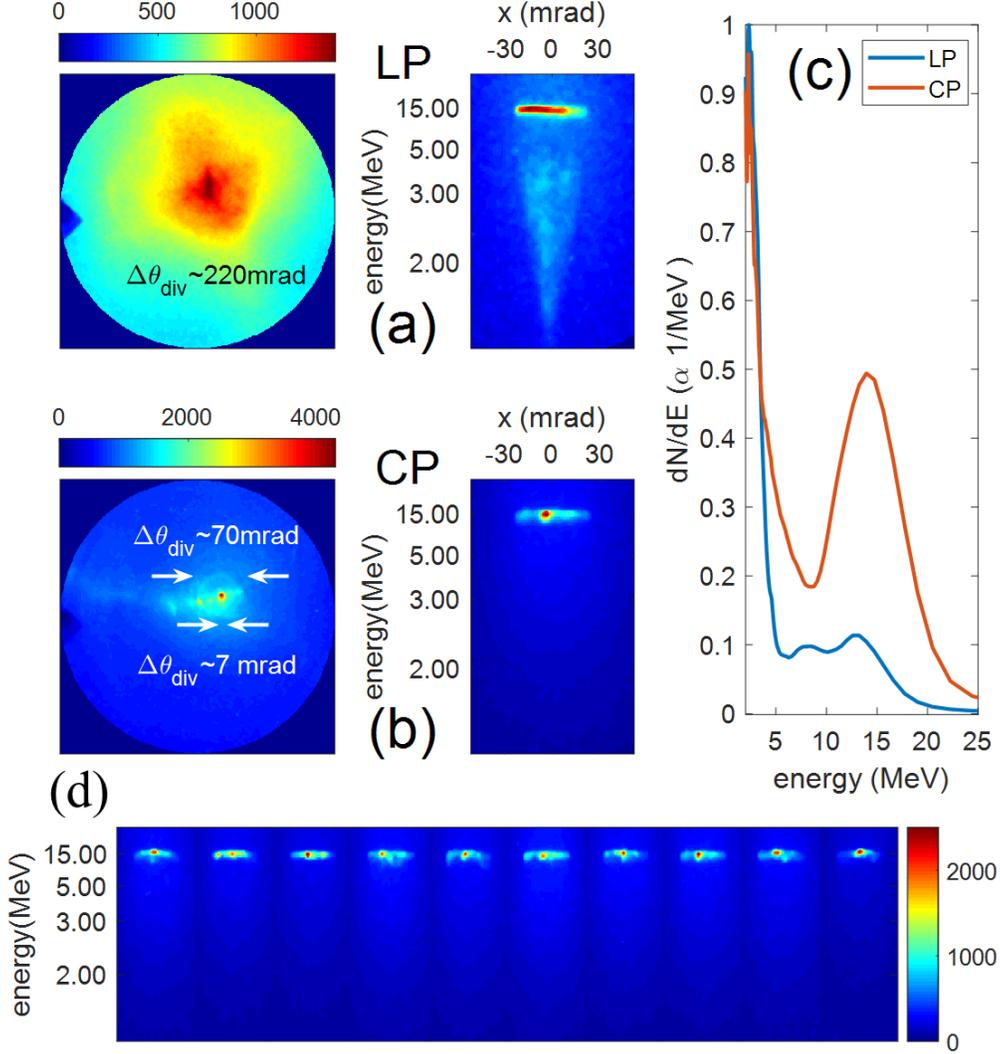

**Figure 6.** 2.7 mJ, 5 fs LP or CP drive pulse, peak jet plasma density $N_{e,peak}/N_{cr} = 0.11$, $(P/P_{cr})_{max} \sim 3.4$, $z_{jet} = 0$. **(a)** LP driver: Electron beam profile (left) and corresponding energy spectrum (right). **(b)** CP driver: Electron beam profile (left) and corresponding energy spectrum (right). The two values of $\Delta\theta_{div}$ shown are for the central peak and the ring-like pedestal. **(c)** Electron spectrum lineouts on linear scale for LP and CP drivers. Electron beam profiles and spectra are integrated over 10 shots. **(d)** Electron spectra for 10 consecutive shots.

The highest electron beam energy for both CP and LP cases occurs when the vacuum beam waist is at the jet center ($z_{jet} = 0$). Figure 6 compares the electron beam profiles and energy spectra for (a) LP vs. (b) CP for pulse energy 2.7 mJ and pulsewidth 5 fs. For CP, the electron beam is quasi-monoenergetic at ~15 MeV with an intense central spot $\theta_{div} < 7$ mrad with < 4 mrad shot-to-shot pointing jitter. This is superimposed on a wider ring-structured pedestal with the same energy, with divergence $\theta_{div}$~70 mrad. Electron spectrum lineouts are shown in Fig. 6(c), and CP-driven beam stability is demonstrated by the spectra of 10 consecutive shots in Fig. 6(d). By contrast, the electron beam divergence of > ~200 mrad for a LP driver is substantially larger. The total detected charge is similar in the two cases, ~2.5 pC. The strong difference in beam divergence is explained as earlier: the few-cycle CP driver pulse mitigates the CEP slip-induced asymmetric oscillation of the wakefield bubble. Our results also show that use of a CP LWFA driver enables stable electron bunches from shot to shot without carrier envelope phase stabilization.

## IV. CONCLUSIONS

We have shown that few-cycle (5 fs), circularly polarized sub-3 mJ laser pulses can generate pC level, low divergence (<10 mrad) electron bunches of energy up to 15 MeV from near-critical density hydrogen jet plasmas. The highest beam energy and lowest bunch divergence was for a circularly polarized drive pulse focused at jet center, with linearly polarized drive pulses producing similar electron energy but order of magnitude larger bunch divergence. We attribute this effect to circular polarization's mitigation of the deleterious effects of carrier envelope phase slip, which induces bubble centroid oscillations. With a circularly polarized driver, the tighter, spiral transverse motion of the bubble centroid leads to less perturbed plasma accelerating fields, and greatly reduced sensitivity to shot-to-shot carrier envelope phase variation. Reduced beam divergence is also achieved by focusing the few-cycle pulse in the plasma density downramp, where both circularly and linearly polarized pulses drive milder wavebreaking and reduced bubble oscillation, with CP pulses driving beams with ~3 × less divergence than LP. Finally, we observe that bunches accelerated in the plasma density downramp have a 2D Lorentzian beam profile consistent with a $\kappa$ ("kappa") electron energy distribution characteristic of a strongly correlated collisionless plasma confined by the quadratic wakefield bubble potential. Kappa distributions had previously been observed only in space or dusty plasmas. We attribute our best fit $\kappa = 2$ to transverse velocity space diffusion [31] imposed on the bunch by red-shifted laser field in the wakefield bubble.

## ACKNOWLEDGMENTS

The authors thank Luke Pascale for technical assistance. This research is supported by the US Department of Energy (DESC0015516), the National Science Foundation (PHY1619582), and the Department of Homeland Security (2016DN077ARI104).

Baudelet, and M. Richardson, *Dramatic Enhancement of Supercontinuum Generation in Elliptically-Polarized Laser Filaments*, Sci. Rep. **6**, 20363 (2016).

**Supplemental material** for "Laser-accelerated, low divergence 15 MeV quasi-monoenergetic electron bunches at 1 kHz"

## 1. Hollow core fiber and chirped mirror compressor

The few-cycle drive beam is generated by broadening the bandwidth of λ=800nm, 35fs FWHM, 6 mJ Ti:Sapphire laser pulses in a helium gas filled, 500µm inner diameter, 2.5 m long hollow core fiber (HCF) by self-phase modulation (SPM). Helium is injected near the fiber exit and pumped out at the entrance. The Ti:Sapphire pulses are focused into the HCF entrance with a 3.3m focal length vacuum telescope to a $2w_0 = 320 \mu m$ ($1/e^2$ intensity) spot diameter. The HCF output pulses are then reflected 8-12 times from two pairs of chirped mirrors (Ultrafast Innovation PC70) and pass through a pair of wedges for fine tuning the group delay dispersion (GDD). Normal incidence reflection off a removable wedge is used for HCF output spectrum measurement, polarization characterization, and pulse length measurement.

A quarterwave-plate (QWP) is used at the HCF entrance to adjust the input pulse polarization. At the HCF exit, a broadband QWP and a Glan polarizer are employed for polarization characterization. For pulse duration measurement, an optical beamline with the same GDD as the main beamline takes a reflection from the removable wedge to a scanning SHG FROG (second harmonic generation frequency resolved optical gating) device [1,2]. The SHG FROG uses a 10µm thick BBO crystal cut for type I SHG at 650 nm to support the full bandwidth. A delay arm, which scans in < 0.1 fs steps, and a stationary arm focus two samples of the beam onto the BBO crystal, and the resulting SHG spectrum is collected by a spectrometer. The full SHG FROG trace is assembled by collecting spectra over a delay arm scan. The uncertainty in pulse length measurement is < 0.4 fs.

The HCF output bandwidth is the broadest, ~200nm FWHM with a double hump shape centered at λ~650nm (see Fig. 1(f) of the main text), when the input pulse is elliptically polarized with $\varepsilon = 0.5$, where $\varepsilon$ is the ratio of minimum to maximum electric field in the polarization ellipse. The measured HCF output polarization in this case is nearly circular (ε~0.9), with a compressed FWHM pulse duration of 5 fs. The HCF output polarization is preserved for the cases with linearly polarized (LP) or circularly polarized (CP) input pulses, but the generated bandwidth in those cases is smaller. For LP input pulses, the bandwidth is slightly narrower than for the optimum elliptical polarization input case ($\varepsilon = 0.5$), with a compressed pulse length of 5 fs. For CP input pulses, the significantly narrower HCF output bandwidth leads to a compressed pulse duration >7 fs (see Fig. 1(f) of the main text) which is ineffective for electron acceleration.

The maximum pulse energy injected into the HCF is limited by self-focusing, leading to ionization of the He gas at the HCF entrance. Significant blue shifting from excessive ionization causes a more complex spectral phase at the HCF output, preventing compression to few-cycle duration. In practice, the maximum input pulse energy is set near the onset of filamentation at the HCF entrance. The final compressed pulses at the target are 5 fs and up to 2.7 mJ. These pulses are focused by a $f/6.5$ off-axis paraboloid (OAP) to a 4.5 μm FWHM intensity spot. The ≳ 50% pulse energy losses are from HCF leakage, the chirped mirror compressor, and the routing mirrors and tuning wedges.

## 2. Electron beam profile and energy spectrum measurement

Accelerated electron bunches are detected using a LANEX fluorescing screen, either for full beam profile measurements, or as the detector in the magnetic spectrometer. A 100 μm thick aluminum foil in front of the LANEX screen blocks laser light and allows passage of high energy electrons. The backside LANEX fluorescence is imaged onto a low noise CCD camera (Andor iXon).

LANEX conversion efficiency is independent of the electron energy for $> \sim 3$ MeV. For lower energy electrons the efficiency becomes energy dependent, but still remains proportional to the total deposited energy. For a given electron bunch energy, the LANEX fluorescence and, in turn, the CCD counts are linear with the incident charge up to the $\sim 100 \text{ pC/mm}^2$ saturation limit [3,4], well above the bunch charges of this paper. For electron beams with a large energy spread, as long as the bunch energy distribution doesn't have a strong transverse dependence, the beam image on the CCD will have the same functional form as the transverse profile of the incident charge. In our experiments, except in the case of electron rings generated in the downramp (see Fig. 4(a), centre panel), there is no transverse dependence of the energy spectrum. For our measured Lorentzian profile beams, there is no transverse spatial variation in the beam energy.

## 3. 2D Lorentzian fits to electron beam profiles

Beam profile and background images are saved in 512×512 matrices. After subtracting background with counts larger than 0.001 of maximum, the resulting beam profile matrix $D(i,j)$ is used for processing. The best 2D Lorentzian fit to $D(i,j)$ is found through minimizing the root mean square (RMS) difference between $D(i,j)$ and a 2D Lorentzian function $L(i,j)$ given by

$$L(i,j) = A[1 + (\frac{x_i - x_0}{w_x})^2 + (\frac{y_j - y_0}{w_y})^2]^{-\kappa + 1/2} \; ,$$

where $(x_0, y_0)$ is the profile center, $w_x$ and $w_y$ are the charge spread in each dimension, and $\kappa$ is the kappa distribution exponent (see main text). The central structure of all beams measured in our experiments is circular or elliptical. In cases where the measured beam profile is elliptical, it is rotated to have major and minor axes parallel to the defined $x$ and $y$ directions in the image coordinate system.

The RMS error is defined as

$$RMS = \left[ \sum_{i=1}^{N} \sum_{j=1}^{N} (D(i,j) - L(i,j))^2 \right]^{\frac{1}{2}} .$$

The parameters of the 2D Lorentzian function are varied in an iterative process using the simplex search method [5] to minimize the RMS error and find the best fit. The normalized error from the best fit is

$$G = (1/N) \left[ \sum_{i=1}^{N} \sum_{j=1}^{N} \left( \hat{D}(i,j) - \hat{L}(i,j) \right)^2 \right]^{\frac{1}{2}} ,$$

where $\hat{D}(i,j) = D(i,j)/D_{max}$ and $\hat{L}(i,j) = L(i,j)/L_{max}$ are the normalized data surface and Lorentzian fit.

## 4. Particle-in-cell (PIC) simulations

Simulations were performed using EPOCH [6], a three-dimensional particle-in-cell simulation code. Simulations were performed with a moving window using a spatial resolution of $\lambda_0/25$ in the axial direction and a resolution of $\lambda_0/12.5$ in the transverse directions, where $\lambda_0 = 650$ nm is the central wavelength of the few-cycle pulse. The density profile was modeled as a Gaussian in the axial direction with a FWHM of 140 μm and peak $H_2$ density and electron density (preionized for Figure 2 and neutral for Figures 4 and 5) corresponding to interferometric measurements of the hydrogen jet target. The number of macroparticles per cell was 25. The laser pulse was modeled as a space and time Gaussian with central wavelength $\lambda_0$=650 nm, no chirp, a FWHM duration of 5 fs, and a FWHM spot size of 4.5 μm. For the CP vs. LP comparison, the pulse was focused from outside of the neutral gas jet. For the other simulations, it was launched at the beam waist from inside the plasma.

## 5. Simulated evolution of > 2 MeV electron beam profiles: CP vs. LP

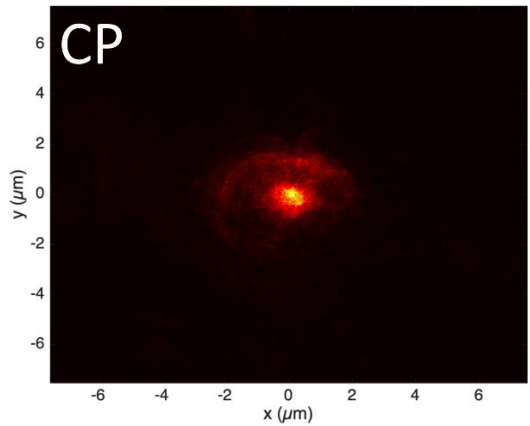

[electron beam profile evolution: CP driver](#)

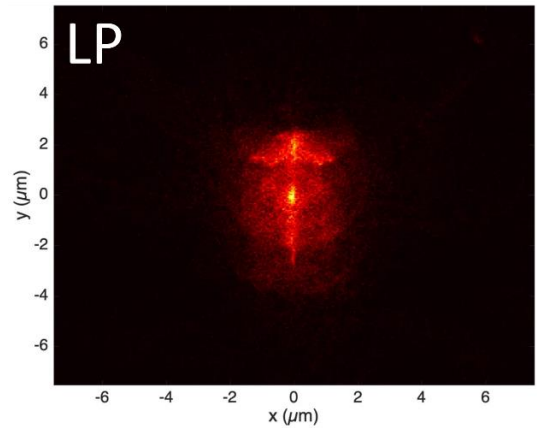

[electron beam profile evolution: LP driver](#)

**Table S1.** Electron beam profiles and evolution movies corresponding to the PIC simulations in Fig. 5 of the main text, for bunch electrons >2 MeV. Top: beam image and movie link for CP driver pulse. Bottom: beam image and movie link for LP driver pulse.